\documentstyle[12pt]{article}

\begin{document}
\baselineskip 18pt
\vspace{.3in}
\title{A New Class of Nonsingular Exact Solutions
for Laplacian Pattern Formation}
\author{Mark B. Mineev--Weinstein and Silvina Ponce Dawson\\[.3in]
Theoretical Division
and Center for Nonlinear Studies, MS--B258\\Los Alamos National Laboratory\\
Los Alamos, NM 87545 }
\maketitle
\begin{abstract}

We present a new class of exact solutions for the so-called
{\it Laplacian Growth Equation}
describing the zero-surface-tension limit of a
variety of 2D pattern formation problems.
Contrary to the common belief, we prove that
these solutions are free of finite-time
 singularities (cusps)  for quite general initial conditions
and may well describe real fingering instabilities. At
long times the interface consists of N separated moving
Saffman-Taylor fingers, with  ``stagnation points'' in between,
 in agreement with numerous  observations.
 This evolution resembles
the N-soliton
solution of classical integrable PDE's.

\end{abstract}
\vspace{.3in}
\hspace{.3in}
\pagebreak

The problem of pattern formation is
one of the most rapidly developing branches of nonlinear science today.[1] Of
special interest is the study of  the front
dynamics between two phases (interface) that arises
in a variety of nonequilibrium
physical systems. If, as it usually happens,  the  motion of the interface
is slow in comparison with the
processes that take place
in the bulk of both phases (such as heat-transfer, diffusion, etc.),
 the scalar field governing the evolution of the interface
is a harmonic
function. It looks natural then, to call the whole process  {\it
Laplacian growth}. Depending on the system,
 this harmonic scalar field is a temperature (in the freezing of a liquid or
Stefan problem),
a concentration   (in solidification from a
supersaturated solution),
an electrostatic potential (in electrodeposition), a
pressure (in flows through porous media), a probability (in diffusion-limited
aggregation), etc.

We present in this paper a new class of solutions of the 2D-Laplacian growth
problem
 in the limit of zero surface tension. These
solutions are quite general because no symmetries of the moving interface
are assumed. Most remarkably,
they do not develop finite time singularities but,
contrary to the common belief,
remain smooth for all finite times. Thus, they may  describe
real fingering instabilities when  surface tension is very small.
In the long-time limit they give rise to N
separated fingers, each of which (for enough separation) describes
the  Saffman-Taylor finger [2] in channel geometry, and
whose
evolution closely resembles the $N$-soliton formation in nonlinear
integrable PDE's.

In the absence of surface tension, whose effect is to
stabilize the short-wave
perturbations of the interface, the problem of 2D Laplacian growth is
described as follows:
\begin{equation}
(\partial^2_x + \partial^2_y)u = 0,
\end{equation}
\begin{equation}
u\vert _{\Gamma(t)} = 0,
\end{equation}
\begin{equation}
\partial_n u\vert _\Sigma = 1,
\end{equation}
\begin{equation}
v_n = -\partial_n u\vert _{\Gamma(t)}.
\end{equation}
Here $u(x,y;t)$ is the scalar field mentioned above, $\Gamma(t)$ is the moving
interface, $\Sigma$ is a fixed external boundary, $\partial_n$ is the
 component of the gradient normal to the interface
(i.e. the normal derivative), and $v_n$ is the normal component of the
 velocity of the front.

We consider in this paper an infinitely long interface. We introduce then a
time-dependent conformal
map $f$ from the lower half
of a ``mathematical'' plane,  $\zeta \equiv \xi + i\eta$,
to the domain of the physical plane, $z\equiv x+iy$,
where the Laplace equation (1) is defined,
$\zeta \buildrel f \over \rightarrow z$.
 We also require that
$ f (t,\zeta) \approx \zeta$ for $\zeta\rightarrow \xi-i\infty$.
 With this definition, the function
$z=f(t,\xi)$  describes the moving interface.

Using this conformal map and taking into account the boundary
conditions of the problem we find:
\begin{equation}
v_n =\frac{{\rm Im}(\bar f_t f_\xi)}{|f_\xi|},
\end{equation}
and
\begin{equation}
- \partial_n u_{\Gamma(t)} = - \partial_l \psi = \frac{\partial \xi}{|\partial
f|} = \frac{1}{|f_\xi|},
\end{equation}
where  the overbar means complex conjugate,
the subscripts $t$ and $\xi$ indicate partial derivatives with
respect to $t$ and $\xi$, respectively,  $\partial_l$ is the component
of  the gradient tangent to
the interface, $\psi$ is a harmonic function of  $x$ and $y$, conjugate
to $u$,  that satisfies $\psi=-\xi$
due to the  boundary conditions (2) and (3).
If we equate Eqs. (5) and (6), in accordance with Eq.(4), we finally obtain:
\begin{equation}
{\rm Im}(\bar f_t f_\xi) = 1
\end{equation}

As in Ref. 3, we will refer to Eq. (7) as  the
{\it Laplacian Growth Equation} (LGE),
because the scalar field determining the growth obeys
the Laplace equation (1).  The
LGE was first derived, to our knowledge, in 1944 independently by
Polubarinova- Kochina[4] and Galin[5]. This equation has remarkable properties,
unexpected  for nonlinear PDE's, such as an infinite set of
conserved quantities[6], a pole decomposition[7], an exact solution in the
class of
finite polynomials[8]  and other exact solutions (though for
particularly symmetrical initial shapes)[9]. Also the Saffman-Taylor finger
[2] is a particular
travelling wave-solution of this equation.
 All these properties (except the latter one)
are nontrivial and nonperturbative due to the
nonlinear nature of the LGE.

Unfortunately, despite  these remarkable properties, practically all known
solutions of the
LGE show  finite-time singularities via the formation of
cusps $[7],[8],[10]$. Therefore,
all these nonperturbative results are helpless to shed light on
 the physics and geometry of the system in the long-term limit.
(Although a few exact results have been presented
that have no finite-time cusps [9],  they correspond to
cases with very restricted
symmetries of the moving interface.)
As a result it has been generally assumed that these finite
time singularities are an essential feature of LGE solutions [11]
and that, in this sense, the physics represented by the LGE is incomplete.
Thus, the natural attitude  was to include surface tension in the theory
 to
stabilize the moving interface
 and get rid of the finite-time singularities.[11]

The main result of this paper is to show that,
contrary to this widespread view,
the LGE (i.e. a zero-surface tension limit of the 2D Laplacian pattern
formation) admits quite a broad class of
exact time-dependent solutions which remain smooth for an infinite time. These
solutions are of ``$N$-finger type'' because they lead to the formation of
$N$ well-developed and separated fingers  in the long-time
asymptotics (see Fig.1). Thus, the problem of finite-time cusps
can be solved within the framework of the LGE. (However, there are problems
in which the inclusion of
surface tension is indeed unavoidable, e.g. the selection
mechanism in dendritic growth).[1]

To introduce this new class of solutions,
we start from the statement that any
function $f(t,\zeta)$ whose derivative, $f_{\zeta}$,
has an arbitrary distribution of moving  poles, $\zeta_k(t)\equiv
\xi_k+i\eta_k$,
and roots, $Z_k(t)$,
 in the upper-half plane,
Im $\zeta>0$, and no other singularities,
 is a solution of the LGE.  This is easy to verify by substitution of any
such function
$f$ into Eq. (7). Let us consider the arbitrary case when  $f_{\zeta}$
has $N+1$ simple poles.
Since we assume that
\begin{equation}
f_\zeta = \prod_{k=1}^{N+1}\frac{\zeta - Z_k(t)}{\zeta - \zeta_k(t)}
\end{equation}
with all
poles  and roots initially lying above the real axis,
we have
\begin{equation}
f = A(t) + \zeta -i \sum_{k=1}^{N+1}\,{\alpha_k\,\log(\zeta - \zeta_k)}
\end{equation}
where $A(t)$ is an unknown function of time, and the factor $-i$ is taken just
for  convenience.

By substitution of the last equation for $\zeta$  real, into the
LGE (Eq. (7)), we find that
\begin{itemize}
\item
 $i)\, \, A(t) = -it \,\,$;
\item
$ii)\, \,  \alpha_k\in{\bf C}$ does not depend on time ;
\item
$iii) $
the time-dependence of the poles $\zeta_k$ is governed by the following
constants of motion $\beta_k$:
\end{itemize}
\begin{equation}
\beta_k \equiv
f(\bar\zeta_k) = -it + \bar\zeta_k -i\sum_{\ell=1}^{N+1} \,\,\alpha_{\ell}\,\,
\log(
\bar\zeta_k - \zeta_{\ell}),\, \, \, 1\leq k\leq N+1.
\end{equation}
Therefore, the motion of the
interface is determined by  Eqs. (9)-(10) for real $\zeta$.
This solution, for $N=1$ corresponds to the development of an isolated
finger, similar to the one found by Saffman for channel geometry. [12]

The break of analyticity of the interface (a cusp) occurs when at least one of
the moving poles, $\zeta_k(t)$, or zeros, $Z_k(t)$,
of  $f_\zeta$  crosses the real
axis, $\eta = 0$, of the mathematical plane, $\zeta$.
If all $\zeta_k$'s
and $Z_k$'s remain on the upper-half plane during the whole evolution, then
 the moving interface remains smooth
(analytic) for an infinite time. To obtain sufficient conditions
under which
 this is true for $f$ given by Eqs. (9)-(10), we note the following:

(i) In order for the solution to exist as $t\rightarrow\infty$
and satisfy $\eta _k >0$ for all finite times,
all $\alpha_k$'s must have positive real part. This is the only way
that the divergent term $-it$ in the R.H.S. of Eq. (10)
can be compensated.
 The term that compensates it is
\begin{equation}
- i{\rm Re}(\alpha_k) \log(\bar\zeta_k - \zeta_k) = -i
{\rm Re}(\alpha_k) \log(-2i\eta_k)
\end{equation}
and implies that
 $\eta_k \rightarrow 0$ as $t\rightarrow\infty$.

(ii) An isolated singularity,
$\zeta_k$, can never reach the real axis at a finite time. If it did,
  the term $-i\alpha_k \log(\bar\zeta_k - \zeta_k)$ in Eq. (10) would diverge
and could not be compensated by any other.
On the other hand, if all Re$\alpha_k$'s are
positive, then groups of $M\leq N+1$ singularities could not reach the
real axis simultaneously at a finite time for exactly the same reason.
Therefore, we can conclude that, if ${\rm Re}\alpha_k>0$ and
$\eta_k(t=0) >0$ for  $1\leq k\leq N+1$, then
$\eta_k(t) >0$ for all finite times $t$ and
 $\eta_k \rightarrow 0$ as $t\rightarrow\infty$.

(iii)We assume first that all $\alpha_k$'s are real
and positive. We calculate the derivative of $f$ with respect to $\zeta
\equiv\xi+i\eta$.
After a little algebra,
we write its
real part
as
\begin{equation}
{\rm Re}\,
  f_\zeta =
1 + \sum_{k=1}^{N+1} \frac{\alpha_k \eta_k}{|\zeta - \zeta_k|^2}
-\eta\sum_{k=1}^{N+1} \frac{\alpha_k }{|\zeta - \zeta_k|^2}
\end{equation}
Since all $\alpha_k$'s and all $\eta_k(t=0)$'s are positive, then by
the result in (ii), also $\eta_k>0$ for all finite times. Thus,
we see from Eq.  (12), that Re$  f_\zeta$
equals zero only if $\eta$ is strictly positive.
This means that the zeros, $Z_k$,
of  $f_\zeta$ lie always on the upper-half of the mathematical
plane and
never cross the real axis.

Consequently, {\it if at time $t=0$,  all $\alpha_k$'s are
real and positive and  all $\eta _k$'s
 are positive, then the interface
represented by Eq.(9) remains  smooth  throughout its evolution}.

For the general case of complex $\alpha_k$'s (with positive real part)
we are currently unable to prove neither the existence nor the
 absence of finite-time
blow-ups. (For $N=0$  we can easily show that the solution
(9) remains smooth for all time).
In spite of the lack of a rigorous proof, we believe that
also in this case,  there are no finite-time cusps for a broad range of
initial conditions.
This conjecture is supported by a series of numerical
experiments that we will discuss in a forthcoming paper. [13]

Eq. (9) is not the only solution of the LGE that is characterized by the
motion  of simple poles. For example, if in Eq. (9)
we replace $\log(\zeta - \zeta_k)$ by $\log(e^{i\zeta} - e^{i\zeta_k})$,
and  introduce a new parameter $\lambda$,
we find a $2\pi$-periodic solution, relevant for channel geometry, of the form:
\begin{equation}
f = -i {{t}\over{\lambda}} +
\lambda\zeta -i \sum_{k=1}^{N}\,{\alpha_k\,\log(e^{i\zeta} -
 e^{i\zeta_k)}}\qquad\qquad -\pi\leq {\rm Re}\, \zeta\leq\pi.
\end{equation}
 This solution must satisfy
$\sum_k\alpha_k = 1-\lambda$ so that
$f_\zeta=1$ for $\zeta\rightarrow -i\infty$. The parameter $\lambda$
is the fraction of width of channel occupied by the fingers.
This solution has the same properties as
Eq. (9).
 In particular, there exist $N$ constants of motion defined by
$\beta_k \equiv
f(\bar\zeta_k)$ and cusps are absent
  if  all $\alpha_k$'s are
real and positive and  all $\eta _k$'s
 are positive, as it follows from the equation:
\begin{equation}
 f_\zeta =
\lambda + \sum_{k=1}^N \frac{\alpha_k }{|1-e^{i(\zeta_k - \zeta)}|^2}
(1-e^{-\eta_k}e^{i(\zeta - \xi_k)})\neq 0\qquad\qquad{\rm if}\, \,
{\rm Im}\, \zeta \leq 0.
\end{equation}
Eq. (13), but with  $e^{-i\zeta}$ instead of $e^{i\zeta}$,
was proposed as a solution of the LGE in [11], where cusps were
found via numerical simulations.
Taking advantage of the corresponding constants of motion, $\beta_k$, we
can easily show the necessity of these cusps if Re$(\alpha_k)>0$
(as is the case in [11], where $\alpha_k=1$ for all $k$).

The constants $\alpha_k$ have a clear geometrical meaning in the physical
plane for both  solutions (9) and (13): $\vert\pi\alpha_k\vert$ is related to
the width of the gap between adjacent moving
fingers (in the case of enough separation), while arg$(\alpha _k)$
is related to the angle that this gap forms with the horizontal.[13]
We show this property in Fig. 1, where we have plotted the
interface $Y\equiv{\rm Im}\, f(t,x)$ vs. $X\equiv{\rm Re}\, f(t,x)$,
 for the solution (9) with two singularities
($N=1$), at a particular time.
The real
parts of both $\alpha$'s are positive, but while  $\alpha_1$
is purely real ($\alpha_1=0.8$), $\alpha_2$
has a nonvanishing imaginary part
(${\alpha_2}=0.8+i 0.1$). We
have drawn on top of each
 gap a dashed line of length  $\vert\pi\alpha_k\vert$
and slope Im$(\alpha_k)/$Re$(\alpha_k)$ that highlights
the meaning of $\alpha_k$.
We conclude then that,
if all  $\alpha_k$'s are real and positive, the interface is a single-valued
function $Y(X)$.
Among the experiments on two-dimensional viscous fingering in a channel, one
can find different degrees of bending and ramification of the moving
fingers.[14] It follows from these experiments that non-single-valued
interfaces generally appear, although with small bending. Therefore,
 complex $\alpha$'s (possibly with small imaginary parts)
would be necessary to describe them using Eq. (9).
As we mentioned before, we expect that even in this more
general setting, Eq. (9) still provides a meaningful (non-singular)
solution, especially, if the imaginary parts of $\alpha$'s are
small enough.

The constants $\beta_k$ also have  a clear geometrical meaning in the
physical plane: the points $({\rm Re}\, \beta _k, {\rm Im}\, \beta_k +
{\rm Re}\, \alpha _k\log 2)$ are
the  coordinates of the tips of the gaps
between fingers.
We show this property  in Fig. 2, where we have plotted the interface
obtained from Eq. (9) at different times
for a particular
choice of $N+1=7$ singularities and real and positive $\alpha _k$'s.
 For the sake of
generality we have deliberately taken the initial condition without
any particular symmetry.
We have indicated in this figure the
location of the tips with asterisks. As one can see,
they are ``stagnation
points'' of the interface. This kind of stagnation points have been
observed in numerous experiments [14] and  numerical simulations [15].
Because all real experiments deal with non-zero surface tension, the
experimental evidence of stagnation points in viscous fingering
supports our conjecture  that a small surface tension
does not destroy the behavior
of the zero-surface-tension solutions. It is interesting
to note that we can also identify
stagnation points occurring in numerical [16] and physical [17] experiments
of
diffusion-limited aggregation (DLA), (a feature usually explained as a
``screening'' effect). This is in accordance with the view (which
we believe) that
DLA fractal growth
is described by the LGE in the continuous limit.[18]

Finally, we show in Fig. 3 the motion of the
singularities in the mathematical plane that correspond to the
solution depicted in Fig. 2. Comparing both figures we can see that,
  while the singularities tend to
the real axis, the interface takes the shape of a set of more and more
well-developed and separated $N$ fingers.
The evolution of this interface resembles the $N$-soliton
solution of  classical exactly solvable PDE's, such as the
Korteweg-de Vries or nonlinear Schr\"odinger  equations,
where in the long-time asymptotics one can also have
$N$ separated solitons, each of which described by
the single soliton solution of the corresponding PDE.
An evident difference is that fingers, unlike ``classical'' solitons,
 always have a non-zero velocity component
normal to the interface.
The connection between the N-soliton
and the N-finger solutions is deeper than a superficial
resemblance, as we will show in
a forthcoming paper.

The separation of the singularities in the long-term asymptotics and
the similarities with the $N$-soliton solutions suggests that there
might be some transformation into action-angle variables in which this
separation arises naturally. We also suspect that some
Hamiltonian structure (at least formal) might underlie the whole evolution.
We are presently working in this direction.
The nontrivial extension of this work
 to the case of a closed
(finite) interface is also of interest. In this case,
our proof regarding the absence of finite-time
singularities
does not hold,
however, the $N$-finger like solution with its associated constants
of motion is still valid. The observation
in physical and numerical experiments of very similar behavior
to the infinite interface, including the existence
of ``stagnation points''[19], makes it seem likely
that the class of solutions studied in this paper may also
shed light on the problem of  radial geometry.

\vskip 0.5cm

We thank  M. Ancona and R. Camassa for their useful comments.
This work is supported by the US Department of Energy at
Los Alamos National Laboratory.

\vfill\eject
\noindent{\bf Figure Captions}
\bigskip

\noindent{\bf Figure 1.} We plot the interface, $Y\equiv$Im $f(t,\xi)$
as a function
of $X\equiv$Re $f(t,\xi)$, at a particular time.
for $f$ defined in Eq. (9) with $A=-it$
and two singularities.
 $\alpha_1=0.8$;
${\alpha_2}=0.8+i 0.1$.
The dashed lines that we have drawn on the gaps between fingers
have  length  $\vert\pi\alpha_k\vert$
and slope Im$(\alpha_k)/$Re$(\alpha_k)$.

\noindent{\bf Figure 2.} We plot the interfaces, $Y$
as a function
of $X$, at times
$t=0$, $t=9$, $t=24$, and $t=30$, for $f$ given by Eq. (9)
 with $A(t)=-it$, $\zeta$ real and $N+1=7$.
The evolution of the seven
singularities was  obtained using a fourth-order Runge-Kutta
integrator to solve the set of evolution equations that are obtained
by differentiating with respect to time Eq. (9) with $A(t)=-it$.
 The paramaters of the simulations are
$\alpha_1 =0.80$, $\beta _1 =6.00-i 37.11$,$\alpha_2=2.00$
$\beta _2 =20.44-i33.98$, $\alpha _3=1.00$, $\beta_3 =35.61-i 34.49$,
$\alpha _4=0.50$,
$\beta _4 =43.15-i 36.31 $,
$\alpha _5 = 1.50$, $\beta _5 = 54.58-i 32.78$, $\alpha_6 =0.35$,
$\beta _6 =64.65-i 40.88$, $\alpha_7 =1.80$, and
$\beta _7 =72.90-i 37.40$.
 We see that, from an initially bell-shaped form, the
interface evolves into a situation characterized by $N=6$
well separated fingers. We also observe the existence of ``stagnation
points'' of the interface (indicated with asterisks).
\medskip

\noindent{\bf Figure 3.} We plot the trajectories
of the moving singularities, $\zeta _k(t)$, on the mathematical plane, that
correspond to the physical process despicted in Fig. 2.
The asterisks indicate, from top to bottom, the points on the trajectory at
times $t=0$, $t=9$, $t=15$,  and $t=30$.
We see that all the singularities go asymptotically
in time to different values and that all $\eta\rightarrow 0$ as
$t\rightarrow\infty$.

\vfill\eject


\begin{thebibliography}{99}
\bibitem{ref18}
It may also be noted that Eq. (9) has an unphysical logarithmic divergence
as
$\vert\zeta\vert\rightarrow\infty$.  The introduction of two
stagnation points far apart from the region of interest, helps us
 avoid this divergence. If the corresponding $\alpha_k$'s
are chosen small enough, then these additional singularities will not
affect the motion of the most relevant ones.
\end{thebibliography}
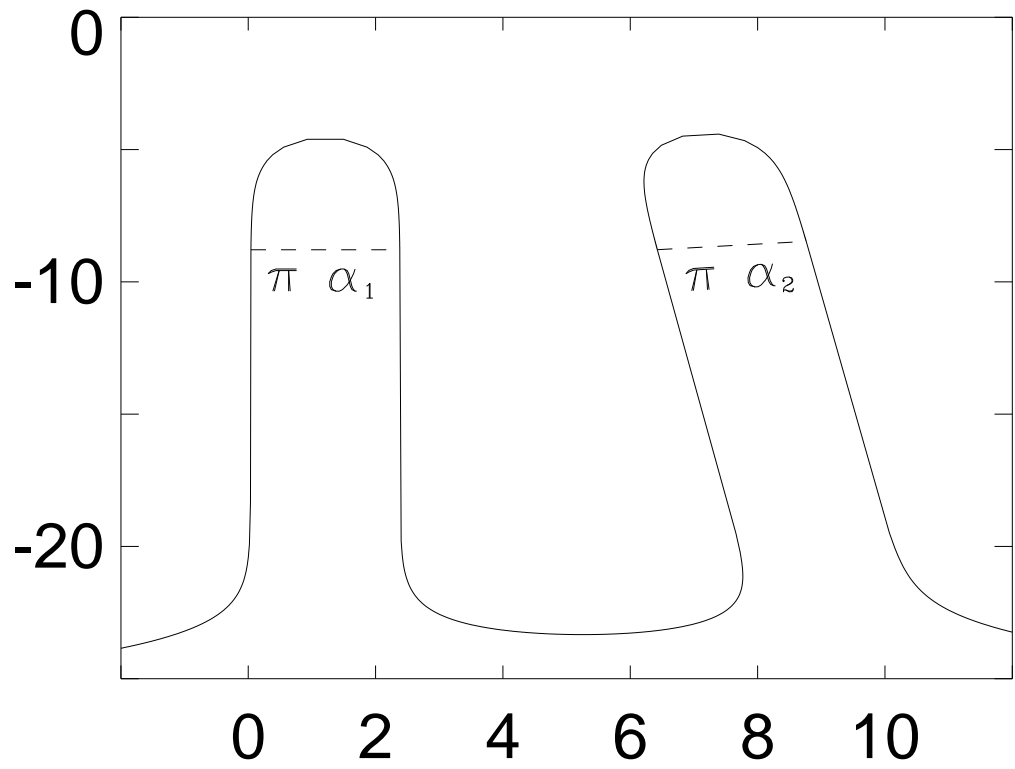
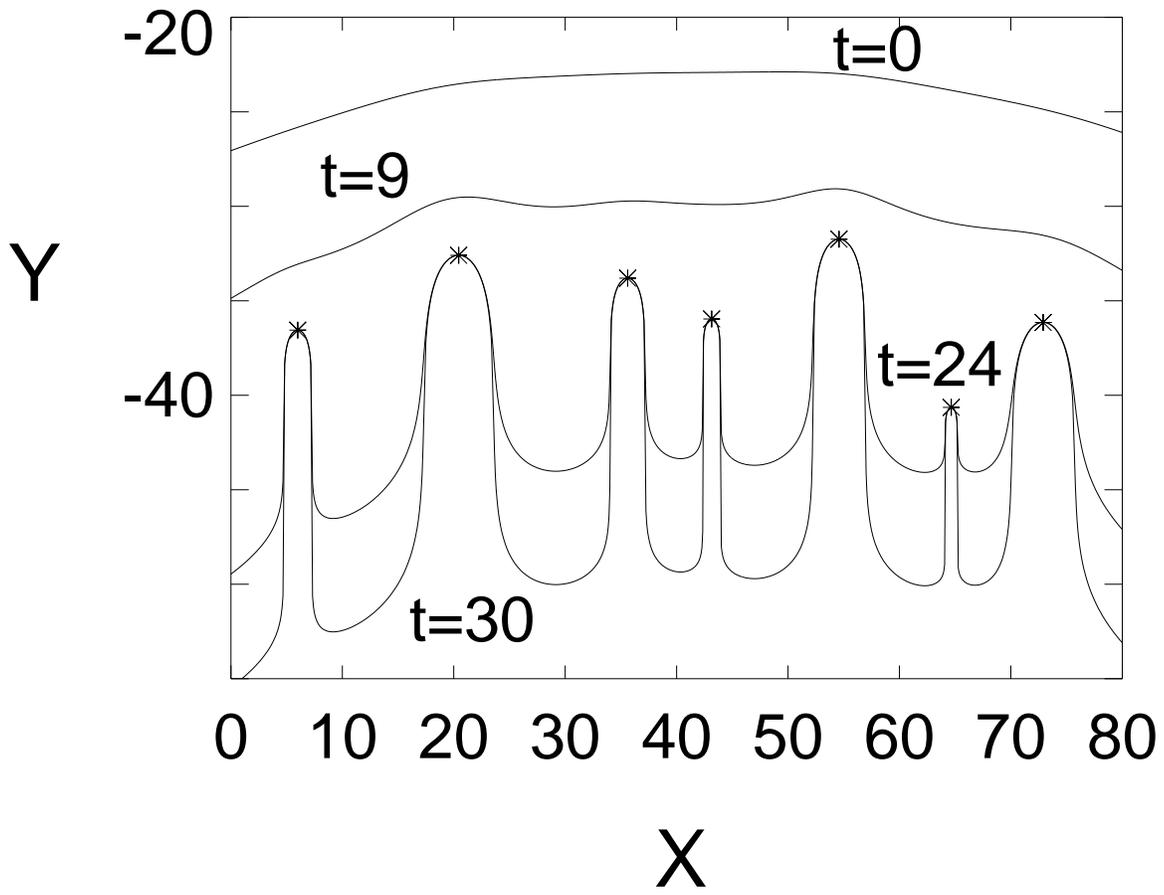
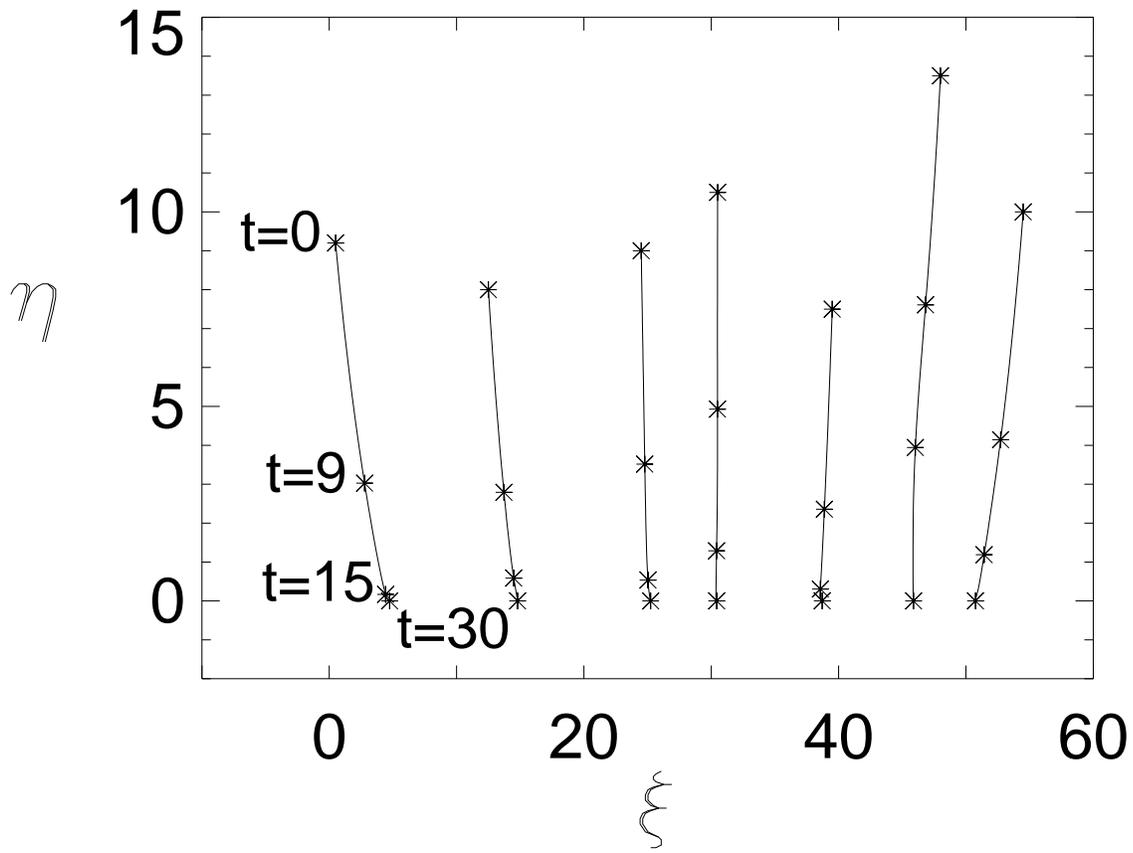
\end{document}